# Prevalence and Major Risk Factors of Non-communicable Diseases: A Machine Learning based Cross-Sectional Study

Mrinmoy Roy
Department of Computer Science,
Northern Illinois University, USA.
mrinmoy.cs10@gmail.com

Anica Tasnim Protity
Department of Biological Sciences,
Northern Illinois University, USA.
protity.microbiology@gmail.com

Srabonti Das
Al Helal specialized Hospital Ltd., Bangladesh
srabontid87@gmail.com

Porarthi Dhar
BGC Trust Medical College, Bangladesh
porarthi.bgc@gmail.com

**Abstract:**

**Objective**: The study aimed to determine the prevalence of several non-communicable diseases (NCD) and analyze risk factors among adult patients seeking nutritional guidance in Dhaka, Bangladesh.

**Participants**: 146 hospitalized adults of both genders aged 18-93 participated in this cross-sectional research.

**Methods**: We collected the demographic and vital information from 146 hospitalized patients in Dhaka, Bangladesh. We checked the physical and vital parameters, including blood sugar, serum creatinine, blood pressure, and the presence or absence of major non-communicable diseases. Then we used descriptive statistical approaches to explore the NCDs prevalence based on gender and age group. Afterward, the relationship between different NCD pairs with their combined effects was analyzed using different hypothesis testing at a 95% confidence level. Finally, the random forest and XGBoost machine learning algorithms are used to predict the comorbidity among the patients with the underlying responsible factors.

**Result**: Our study observed the relationships between gender, age groups, obesity, and NCDs (DM, CKD, IBS, CVD, CRD, thyroid). The most frequently reported NCD was cardiovascular issues (CVD), which was present in 83.56% of all participants. CVD was more common in male participants. Consequently, male participants had a higher blood pressure distribution than females. Diabetes mellitus (DM), on the other hand, did not have a gender-based inclination. Both CVD and DM had an age-based progression. Our study showed that chronic respiratory illness was more frequent in middle-aged participants than in younger or elderly individuals. Based on the data, every one in five hospitalized patients was obese. We analyzed the co-morbidities and found that 31.5% of the population has only one NCD, 30.1% has two NCDs, and 38.3% has more than two NCDs. Besides, 86.25% of all diabetic patients had cardiovascular issues. All thyroid patients in our study had CVD. Using a t-test, we found a relationship between CKD and thyroid (p-value 0.061). Males under 35 years have a statistically significant relationship between thyroid and chronic respiratory diseases (p-value 0.018). We also found an association between DM and CKD among patients over 65 (p-value 0.038). Moreover, there has been a statistically significant relationship between CKD and Thyroid ($P < 0.05$) for those below 35 and 35-65. We used a two-way ANOVA test to find the statistically significant interaction of heart issues and chronic respiratory illness, in combination with diabetes. The combination of DM and RTI also affected CKD in male patients over 65 years old. Among machine learning algorithms, XGBoost produced the highest accuracy 69.7% in comorbidity detection. Random forest feature importance detected age, weight and waist hip ratio as the major risk factors behind the comorbidity.

**Conclusion**: The prevalence study helps to identify the future risks and most vulnerable groups. By initiating and implementing control plans based on the prevalence study, it is possible to reduce the burden of NCDs on the elderly and middle-aged population of Bangladesh.

**Introduction:**

Non-communicable diseases are the ones that do not have a clear etiology, are not contagious, and are chronic. Once they occur, there is almost no panacea for an absolute cure. Often called lifestyle diseases, non-communicable diseases are generally associated with heredity, diet, stress, environmental exposure, and socioeconomic factors. Even 50 years ago, the average life expectancy of a human being was 30 years. However, the discovery, implementation, and proper administration of immunization and antimicrobial drugs effectively reduced infectious diseases by successfully eliminating and preventing major outbreaks and epidemics. All these will likely increase the average human lifespan to 73 years in 2022 [1]. Nevertheless, this triumph over infectious diseases cannot prevent humans from becoming sick. Instead, we see a gradual increase in non-communicable diseases (NCDs) worldwide. This alarming paradigm shift from communicable to non-communicable disease is affecting millions from different parts of the world. Today, NCDs are the leading cause of global morbidity, disability, and disease burden. Heart disease, diabetes, stroke, chronic lung diseases, and cancer are responsible for 74% of mortality worldwide (41 million deaths yearly) [2]. 77% of this NCD-related morbidity occurs in low and middle-income countries [3]. In 2019, NCDs killed 1.62 billion people, whereas infectious diseases killed only 13.7 million [4, 5]. Increased age, physical inactivity, unhealthy diet associated with high fat and high salt intake, low fruit and vegetable consumption, obesity, occupational stress, and tobacco or alcohol consumption are some of the risk factors for these diseases [6]. Elevated blood sugar (hyperglycemia), high blood pressure (hypertension), high cholesterol (hyperlipidemia), and obesity are some of the physiological parameters of NCDs. As the average lifespan increases globally, the number of elderly individuals with multiple chronic NCDs is expected to rise, too. However, NCDs not only cause elderly morbidities but also affect people under 70, causing the premature death of 17 million people worldwide [2].

South Asia is one of the most densely populated areas of the world. People from South Asia suffer from poverty, limited healthcare access, illiteracy, overnutrition, and malnutrition issues. According to [7], about 50% of all South Asian adult diseases are NCDs. Like other low-income countries, Bangladesh faces all these issues while undergoing an epidemiological transition from infection-related morbidity to non-infectious morbidity. Thirty years ago, the percentages of death caused by communicable and non-communicable were 8% and 52%, respectively [7]. These percentages were 11% communicable and 68% non-communicable diseases in 2006. In 2019, NCDs accounted for 70% of Bangladesh's total deaths [8]. A national-level study performed on 18-69 years old individuals living in rural and urban areas in 2019 showed that 70.9% participants had at least one or two NCD risk factors [4]. NCDs were frequently observed in rural settings, which is an alarming concern [9]. The urban slum residents, too, were found to have certain risk factors for NCDs, such as frequent tobacco use, diabetes, higher lipid profiles, overweight, underweight, low fruit and vegetable uptake, sedentary lifestyle, and hypertension [10, 11]. Moreover, the post-covid era worsened healthcare accession, treatment, and follow-ups, affecting overall NCD conditions. In another study conducted with older adults aged between 60-69 years living in rural areas, 58.9% had at least one NCD, and 22.9% had co-morbidities [12]. Not only the elderly adults are affected, but the younger generations are also prone to develop NCDs too. According to another study, almost all Bangladeshi teenagers aged from 12 to 17 years had at least one risk factor for NCDs [13] and the university postgraduate students with a mean age of 26 had a high proportion of risk factors for NCDs [14].

Diabetes mellitus (DM), cardiovascular diseases (CVD), chronic kidney diseases (CKD), chronic respiratory diseases (chronic obstructive pulmonary disease and asthma), and cancers are the leading NCDs. Previous reports suggested that diabetes and cardiovascular disease are the leading NCDs in Bangladesh [15, 16, 17]. In 2021, more than 13 million people had diabetes, with an estimated 75,617 deaths related to it [15]. Every 1 in 5 Bangladeshi people has hypertension, suggesting a high number of cardiovascular cases over the decades [16]. In 2018, the overall prevalence of CVD in Bangladesh was 21% [17]. The CKD prevalence in 2019 was 22.48%, higher than the global CKD prevalence (34). According to Bank et al., in Dhaka, the prevalence of chronic obstructive pulmonary disease (COPD) was 11.4% in 2013 [18]. Irritable bowel syndrome (IBS) and thyroid dysfunction are two less frequent non-communicable diseases. Research shows that the prevalence of IBS was 7.2% among the Bangladeshi population in 2017 [19]. A limited study was performed on thyroid dysfunction data. In one study conducted in 2010, the prevalence of thyroid dysfunction was 10.03% [20].

Gender is a key factor in terms of health. Men's health varies substantially from women's due to the difference in body structure and behaviors. Though men are bigger and more muscular than women, it shows a different history in medical terms. According to National Center for Health Statistics, the gender gap in life expectancy increases for females in America from 2 to 5 years over the 1900 to 2017 timeline [21]. More than 39% of men have heart issues after age 65 compared to about 27% of women of the same age group [22]. Men don't have the protection of estrogen, which keeps women's cholesterol level in check and reduce heart disease risk factor. Even in their twenties, women are far less likely to have hypertension (12%) compared to men (27%) [23]. Moreover, type-2 diabetics are higher in men, especially at the ages of 35-54, but the complications are severe for women after being affected. Men are more prone to develop diabetes due to their uncontrolled lifestyle and inactivity, insulin resistance, weight gain, smoking habits, and alcohol consumption [24]. In the case of chronic kidney disease (CKD), though the prevalence of CKD is higher in women, the progression of CKD to end-stage kidney disease (ESRD) is 50% higher in adult men [25]. Moreover, Diarrhea predominant IBS (IBS-D) is more common in men (44.8%), whereas IBS with constipation and IBS with alternating bowel habits are more (39.3%, 42.6% respectively) found in women [26], and the severity of IBS symptoms, increased inflammatory cytokines impaired female quality of life more than male. The incidence of respiratory symptoms is also higher (230/1000 patients per year) in female patients which is 180/1000 in male patients per year [27]. 84 studies from the PubMed database provided information that females are more commonly affected with infections of the upper respiratory tract (RTI) and males in the lower respiratory tract. Most RTIs, especially community-acquired pneumonia, are severe in males leading to a higher mortality rate [28]. However, the prevalence of thyroid disease was found to be much higher in females than males, approximately 1 in 8 women will be affected by thyroid condition at some point in their lives [29]. According to Bauer et al., 20% of postmenopausal women are found to have subclinical hypothyroidism and this disturbance in thyroid system function complicates the diagnosis and treatment of mood and cognition disorder [30].

This paper is based on cross-sectional demographic data from 146 patients of both genders, aged from 18 to 95 years old, seeking one-on-one nutritional counseling. The patients had been admitted to a renowned hospital in Dhaka, Bangladesh, for a specific period. The demographic information about their age, sex, height, weight, waist and hip measurement, and vital statistics, including blood sugar, serum creatinine, blood pressure, and the presence or absence of NCDs, were recorded. Based on this information, this study provided insight into the prevalence of four leading NCDs, diabetes mellitus, chronic kidney disease, cardiovascular disease, chronic respiratory illness, and two less-studied NCDs, irritable bowel syndrome and thyroid dysfunction. After determining their overall prevalence percentage, we calculated any possible correlation between NCDs and other factors, such as age, gender, hypertension, and obesity. The correlation of different NCDs is discussed based on statistical analysis. We used a machine learning algorithm to predict the comorbidity based on patients' demographic and physical information along with their feature importance. We also discussed the effects of urbanization, socioeconomic factors, environmental pollution, and climate change that might affect this NCD prevalence and suggested preventive approaches for managing NCDs.

**Ethics:**

This study follows all the ethical considerations as per the guidelines of the Bangladesh Medical Research Council (BMRC) and World Health Organization (WHO). Data collection permission was taken from the administrative office of the respective hospital. Each participating patients of the hospital gave their oral consent before the data collection phase. We describe the objectives and procedures of the research to the participants and also their right to refuse and withdraw at any stage of the study.

**Data and Study Population:**

A hospital-based cross-sectional study was performed to estimate the prevalence of NCDs. Data was collected from a 150-bed hospital (located in the north part of Dhaka city corporation, Bangladesh. The hospital is selected for its low to high socio-economic status patients. It also provides primary to tertiary health care which makes the study population diversify. The cross-sectional study was carried out over a period of 3 months from October 2022 to December 2022 on the continuously monitored hospitalized patients who opted for diet plan from nutritionists during their stay. The study participants were previously informed and upon their consent with the help of their nutritionist, NCD related data were collected from the individual patient file. The patient file includes important past information about the patients such as surgeries, accidents, hospitalizations, and their recent diagnostics test results.

A total of 172 patients' data were gathered during the study period. However, due to some missing information in patients' files, 146 patients were selected at the end. Among them, 86 were male and 60 were female. We tried to gather information on patients of various ages, from 18 to 95, and in mild to severe conditions. Among different NCDs, Diabetes, heart disease, chronic kidney disease, choric respiratory disease, irritable bowel syndrome, and thyroid problems are considered for the prevalence study because of their frequency in a substantial amount of the Bangladeshi population. Bangladesh is ranked $8^{th}$ by IDF due to the highest number of adults (20-79 years) with diabetes [31]. Additionally, in the last 10 years, cardiac arrest related death has increased 35 times for men and 48 times for women in Bangladesh [32]. High levels of pollution in the air and unhealthy lifestyle increased the choric respiratory disease rate in the last few years. The CKD prevalence in Bangladesh is 22.48% which is more than the global CKD prevalence [18]. Moreover, IBS is a major health issue in both urban and rural population in Bangladesh [19] and it is common to have thyroid dysfunction with diabetes mellitus [33]. That's why, based on the prevalence and their life-threatening consequences on the overall population, we selected these NCDs in our study.

However, we also collected patients' demographic information and performed physical measurements to get their physical attributes: height, weight, and waist-hip ratio. Digital weighing scales were used to determine the weight of a patient due to its accuracy. A stadiometer was placed against a straight wall with no baseboard or unevenness to ensure accurate height readings. The patients to be measured stood below the height meter, barefoot with relaxed shoulders and the scapula, buttocks, and heels touching the wall.

Waist circumference was measured by wrapping the measurement tape around the narrowest part of the stomach, near or just above the belly button. While the tape rested gently on the skin, the measurement was taken on the exhalation. Hip measurement was taken around the widest part of the hips and buttocks. After that, we got the waist-hip ratio by dividing the waist size by hip size.

Table 1: Study Variables

| Variables | n | Data Type | Mean | Minimum | Maximum | Standard Deviation | 50th Percentile |
|---|---|---|---|---|---|---|---|
| Gender | | | | | | | |
| Male | 86 | Categorical | - | - | - | - | - |
| Female | 60 | Categorical | - | - | - | - | - |
| Age Group | | | | | | | |
| <=35 | 15 | Categorical | - | - | - | - | - |
| 35<>65 | 67 | Categorical | - | - | - | - | - |
| >=65 | 64 | Categorical | - | - | - | - | - |
| Age | - | Numeric | 59 | 18 | 95 | 16.43 | 60 |
| Weight | - | Numeric | 62.2 | 90 | 33 | 12.73 | 60 |
| Height | - | Numeric | 162.26 | 124.46 | 175.26 | 9.49 | 165.1 |
| Waist Hip Ratio | - | Numeric | 0.76 | 0.47 | 0.99 | 0.12 | 0.76 |
| DM | 80 | Numeric | 10.56 | 3.4 | 25.5 | 4.39 | 9.90 |
| CKD | 42 | Numeric | 2.54 | 1.20 | 5.40 | 1.01 | 2.35 |
| Heart Issue | 122 | Categorical | - | - | - | - | - |
| IBS | 2 | Categorical | - | - | - | - | - |
| RTI | 51 | Categorical | - | - | - | - | - |
| Thyroid | 12 | Categorical | - | - | - | - | - |

**Study Variables:**

The variables considered for this study are patients' demographic and physical information with their underlying diseases. We collected each patient's gender, age, height, weight, waist-hip ratio as physical and demographic information shown in table 1. Height was measured in cm, and weight was measured in kilograms. For calculating the waist-hip ratio variable, waist and hip both are measured in inches. Regarding DM, fasting blood glucose test values below 5.7 mmol/L and above 3.9 mmol/L were considered normal and coded as "No". The test value above 5.7 mmol/L was considered hyperglycemia and coded as 'High'. In contrast, the values below 3.9 mmol/L were called hypoglycemia and input as 'low''. Serum creatinine value was used to detect CKD. A creatinine level greater than 1.2 mg/dL for women and greater than 1.4 mg/dL for adult men was considered as CKD [34]. We checked patient history to find any heart related diseases and marked ''Yes'' for heart problems and 'No' otherwise.

Patient blood pressure was measured using a sphygmomanometer because of its better accuracy than a digital device [35]. Blood pressure was measured in millimeters of mercury (mmHg). Diastolic blood pressure less than 60 mm Hg and systolic blood pressure less than 90 mm Hg [36] were considered hypotension and marked as 'low'. Diastolic pressure 70-80 mmHg and systolic pressure 90-120 mmHg were marked as 'normal' and more than these values were considered hypertension, marked by 'high'. Based on patients' experience, previous history, and continuous monitoring, IBS, respiratory illness, and thyroid issues were marked down. If any of these diseases were found in a patient, ''Yes'' was used under the corresponding disease and ''No'' otherwise.

**Statistical Analysis:**

The data were analyzed and visualized using different packages of python. With standard packages of NumPy, pandas, matplotlib, we used statistical function SciPy.Stats for different hypothesis testing. However, there always exists some level of uncertainties in hypothetical testing as tests are based on probability. There are mainly 2 types error: type I and type II. Type I means wrongful assumption that hypothesis testing has worked even though it hasn't, which is called false positive. Type II error occurred when null hypothesis is false, but the test subsequently fails to reject it. This error is also referred as false negative error, which is the opposite of Type I error. These errors are very crucial in our case because rejecting a disease prevalence test despite being true and accepting a null hypothesis which is false will not reflect the proper scenario of NCDs prevalence in patients. Therefore, for balancing these two errors, we selected the threshold of statistical significance, $\alpha$ as low as 0.05 so that the confidence interval becomes 95% and there is only 5% chance of making errors.

Firstly, we described the dataset using descriptive statistics: minimum, maximum, average, $25^{th}$ percentile, $50^{th}$ percentile, $75^{th}$ percentile, standard deviation, and frequency distribution. A one-sample proportion test was conducted to ensure the sample proportion of gender includes the population proportion. Then each disease prevalence, based on gender, along with their percentage of the total affected population, was calculated. We also calculated the percentage of disease prevalence in different age groups to find out the age group-based disease frequency. We also analyzed the disease prevalence rate based on the combination of gender, age group, and waist-hip ratio.

The Chi-Square test was applied to determine the association between gender and diseases. We also explored the relationship between age groups and diseases. T-test was used to compare the sample mean of systolic and diastolic blood pressure with their respective hypothesized value (according to CDC) and to find any significant difference. Moreover, a one-tailed T-test was performed to detect the direction of differences. We also determined whether there is any statistically significant relation between the numerical attributes (age, waist-hip ratio) and NCDs using T-test. The combination effect of diseases on another disease was also analyzed using a two-way ANOVA test. Moreover, we used the random forest machine learning algorithm to predict the number of diseases in a patient. We also leveraged random forest feature importance to find the relevant features in the classification task.

**Results:**

**Demographic study: age, sex, height, and weight:**

In our study, we collected demographic information from 146 diet-seeking patients ranging from 18 years to 95 years of age. Of them, only 15 (10%) were below 35, 67 (46%) were between 35 to 65, and 64 (44%) were 65 and above. The mean age was 59 (95% CI = 24.14, 91.86), with a standard deviation of 16.43. The $25^{th}$, $50^{th}$, and $75^{th}$ percentile fell within the 48.5, 60, and 72 years of age range. Of the 146 participants, 86 were male, and 60 were female, which gave us approximately a 60:40 male: female ratio (Figure 1).

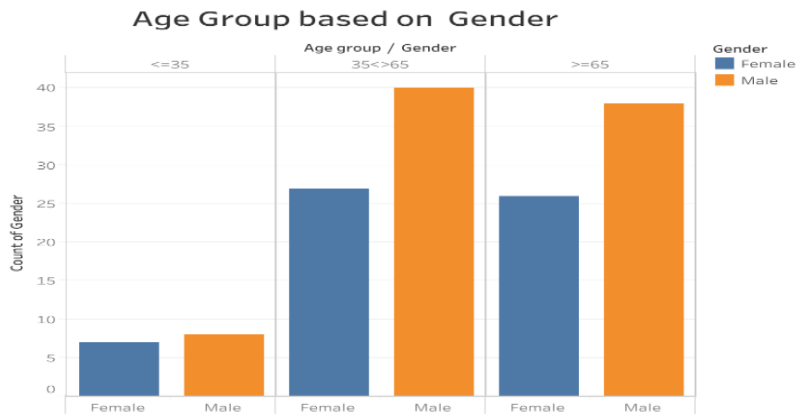

Figure 1. Gender-based Age Group Distribution of Participants

The overall height ranges from 124.46 cm to 175.26 cm, irrespective of gender (Male: 152.4 cm–- 175.26 cm, Female: 124.46 cm -170.18 cm) The mean height value was 162.26 cm (95% CI = 143.28, 181.24), with a standard deviation of 9.49, and a $75^{th}$ percentile value of 170.18 cm. The weight ranges from a minimum of 33 kg (72 lbs.) to a maximum of 90 kg (198 lbs.) (Male: 40 kg (88.18 lbs.)– 90 kg (198.46 lbs.), Female: 33 kg (72.75 lbs.) – 86 kg (189.59 lbs.)) with a waist-hip ratio from 0.47 to 0.99 (Male: 0.49 – 0.99, Female: 0.47 – 0.97).

## Analysis of blood pressure:

We observed the blood pressure of all individuals and found out that hypertension is not only affecting elderly individuals but also among people between 35-65 years of range. Of our 146 participants, 53.4% had hypertension, and a few had low blood pressure. Both genders have hypertension, but the frequency is more significant among males. Of them, 60.2% were men, and 39.7% were women (Figures 2).

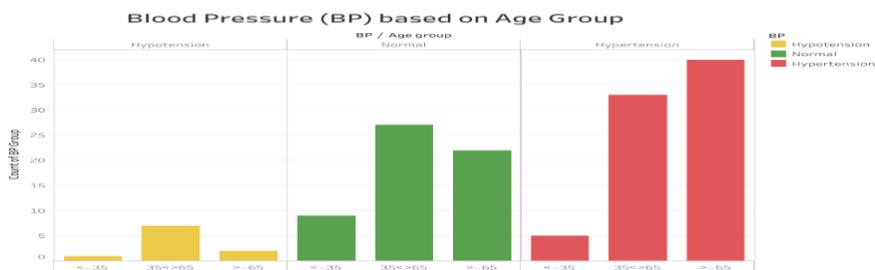

Figure 2. Age-based Distribution of Hypertension

## Diabetes mellitus (DM):

In our study, out of our 146 participants, 54.79% were diabetic. The blood sugar value of our patients ranged from 61 mg/dL (3.4 mmol/L) to 459.45 mg/dL (25.5 mmol/L), indicating that the diet-seeking patients have developed either hypoglycemia or hyperglycemia, depending on the overall health condition (Figure 3). Of the total diabetic patients, 52.5% were men, and 47.5% were women. In our data, diabetes is frequently observed in both genders.

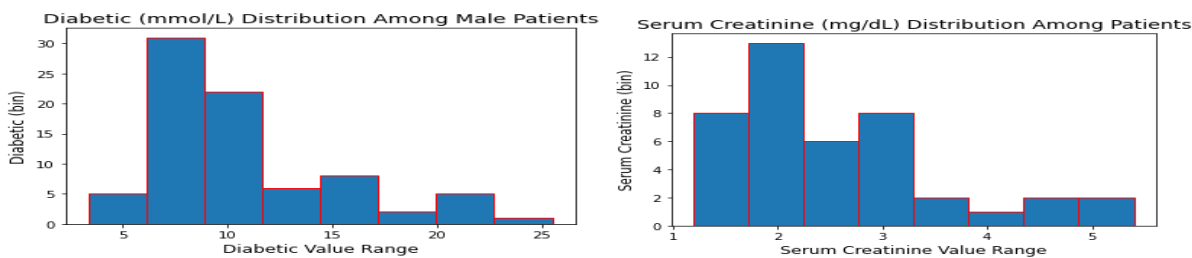

Figure 3. DM Distribution (left) and Serum Creatinine Distribution (right) among patients.

## Chronic kidney diseases (CKD):

Our study reported that the CKD patients had serum creatinine levels ranging from 1.2 mg/dL to 5.4 mg/dL. Of 146 individuals, 28.77% were associated with CKD (Figure 3). Of the total CKD patients, 54.76% were male, and 45.24% were female.

## Irritable Bowel Syndrome (IBS):

In this study, only 1.37% of clients had irritable bowel syndrome. All of them were female. 1.25% of female IBS patients were under 35 years old, and the rest, 1.25%, were greater than 65. 50% of IBS patients had Diabetes mellitus, CKD, chronic respiratory illness, and cardiovascular disease. We could not retrieve much information on this disease with this limited data.

**Cardiovascular disease (CVD):**

The most frequently observed NCD in this study was cardiovascular illness. 122 of 146 patients had cardiovascular illnesses. This percentage (83.56%) is the highest of all NCDs. 62.3% of them were males, and 37.7% were females. 12.5% were under 35, 67.5% fell between 35 and 65, and 72.5% were above 65. In this case, CVD had both gender and age-based predisposing factors.

**Chronic respiratory diseases (CRD):**

Our study shows that 34.93% of total patients had chronic respiratory diseases. 52.94% of the chronic respiratory patients were male, and 47.06% were female. This disease is almost equal in both genders. However, 3.75%, 32.5%, and 27.5% of individuals with respiratory illness fell into the <35, 35-65, and >65 years age groups, respectively.

**Thyroid issue:**

In our study, 8.22% of individuals had thyroid issues. Of them, 58.33% were male, and 41.67% were female. Only 1.25% of people were under 35 years old. 7.5% were middle-aged (35-55 years old), and 6.25% were 65 and above.

**NCD distribution over age and gender:**

Age and gender are the two most common risk factors for progressive NCDs. Among 146 total hospitalized individuals who sought nutritional counseling, only 10.27 % were under 35 years old. 45.89 % fell into the 35-65 age group, and 43.84 % of clients were 65 and above. Here we can see that almost 90% of diabetic individuals are middle-aged, aged 35 and above. 3.75% of the total diabetic females were less than 35 years old. 21.25% of them fell into the 35-65 years age group. The rest (22.5%) were 65 and above. Only 1 male (1.25%) under 35 had diabetes. 25% of the total diabetic male were between 35 to 65 years. The rest (26.25%) were greater than 65 years old. Here, we can see that gender does not play any predisposing role in diabetes prediction. Instead, this disease is associated with age. 95% of diabetic individuals are over 35, irrespective of sex (Figure 4).

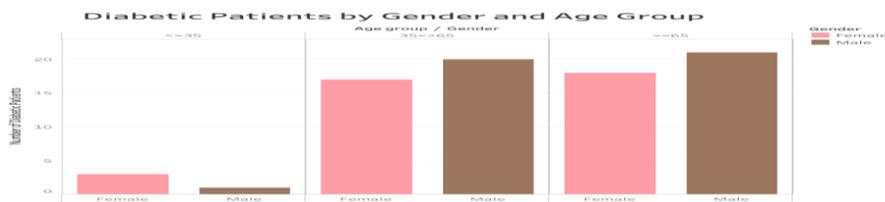

Figure 4: Age and Gender-based DM Distribution.

In terms of CKD, no female under 35 was diagnosed with CKD. On the other hand, 2 males under 35 years (2.5% of the total CKD population) developed CKD. 14.29% male and 14.29% female CKD patients aged between 35-65 years. 65 and above individuals were the worst sufferer of CKD, with 35.71% being male and 30.95% being female. Like DM, CKD also has an age-based progression (Figure 5).

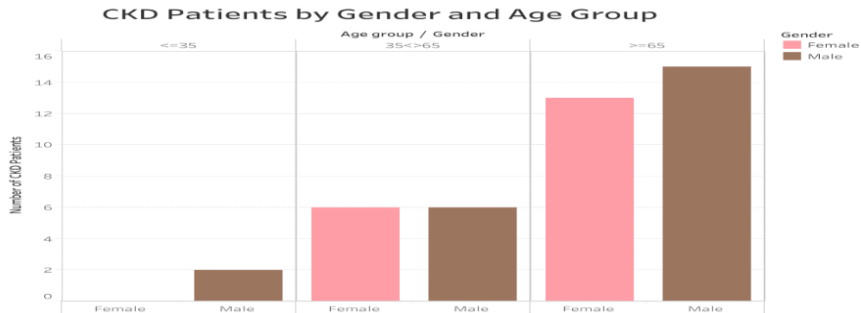

Figure 5: Age and Gender-based CKD Distribution.

CVD was associated with senescence and gender, affecting more older individuals than younger ones and more men than women. 1.64% of CVD patients were female under 35, 17.21% were female of 35-65, and 18.85% were female of 65 and above. 6.56% of CVD males under 35, 27.05% of males aged 35-65, and 28.69% of males aged 65 and above were diagnosed with heart-related issues (Figure 6).

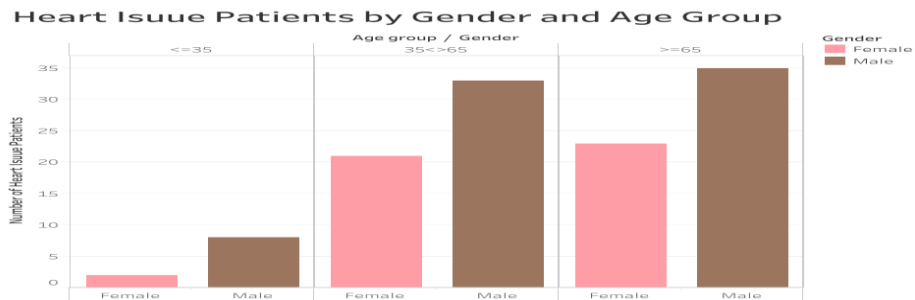

Figure 6: Age and Gender-based CVD Distribution.

Of all patients who visited for nutritional counseling, 20%, 38.81%, and 34.38% of individuals with respiratory illness (CRD) fell into the <35, 35-65, and >65 years age categories, respectively. 3.92% of the total chronic respiratory patients were females under 35, 19.61% were 35-65 years old females, and 23.53% were females older than 65. 1.96% of the total chronic respiratory patients were males under 35, 31.37% were 35-65 years old females, and 19.61% were males older than 65. Here we can see that chronic respiratory illness is more prominent in the middle-aged population than in younger or older ones.

Out of 146 patients, 6.67% of individuals under 35, 8.96% between 35-65, and 7.81% of 65 and above had thyroid-related issues. No female under 35 had thyroid illness. 33.33% of females in the 35-65 age range and 8.33% of females of >65 years had thyroid. 8.33% of males under 35, 16.67% of males in 35-65 age range, and 33.33% of males of >65 years had thyroid.

**NCD distribution over waist-hip ratio (WHR):**

Based on the waist-hip ratio of all participants, we have found 28 (19.17%) obese individuals. Of them, 12 (8.2%) were male with a WHR $\geq$ 0.9, and 16 (10.9%) were female with a WHR $\geq$ 0.85. Diabetic individuals have a waist-hip-ratio range of 0.55-0.96 and 0.48-0.96 for men and women, respectively. People having CKD had a waist-hip ratio ranging from 0.49 to 0.96 for males and 0.55 to 0.96 for females. The waist-hip ratio of female IBS patients fell between 0.65 and 0.8. The waist-hip ratio for CVD patients fell between 0.49-0.99 for males and 0.48-0.97 for females. The male and female waist-hip ratios for the CRD participants were 0.49-0.99 and 0.48-0.96, respectively. For patients having thyroid issues, the waist-hip ratio range fell between 0.49-.096 for males and 0.68-0.84 for females.

**An interaction-based study between different NCDs:**

Interactions were observed with different NCDs. For example, 36.25% of all diabetic patients developed chronic kidney diseases. 17.5% and 18.75% of them are male and female, respectively. Only 1 diabetic person had IBS. Interestingly, our findings showed more correlation between DM and CVD. 69 out of 80 diabetic individuals, or 86.25% of all diabetic patients, including 37.5% females and 48.75% males, had heart-related conditions to some extent. 36.25% of the total diabetic patients had chronic respiratory illnesses. Of them, 15% were males, and 21.25% were females. Only 8.75% of diabetic individuals have thyroid issues, with the female and male percentages being 5 and 3.75, respectively.

69.05% of CKD patients (33.33% men and 35.71% women) also had diabetes mellitus. 2.38% of CKD patients had IBS. 90.48% of CKD patients had developed cardiovascular illness. Of them, 42.86% were females, and 47.62% were males. 35.71% (16.67% of males and 19.05% of females) of CKD patients had chronic respiratory illnesses. 16.67% (7.14% of males and 9.52% of females) of CKD individuals had thyroid issues, too.

56.56% of cardiovascular patients (31.97% males and 24.59% females) had DM. only 0.82% of them had IBS. 31.15% of the patients (16.39% males and 14.75% females) had CKD. 36.89% of CVD patients had chronic respiratory disease. Of them, 19.67% were males, and 17.21% were females. 9.84% of CVD individuals (5.74% males and 4.1% females) had thyroid disease. We have a more significant association between CVD and DM among the patients.

56.56% of all CRD patients (23.53% males and 33.33% females) had DM. Only 1.96% of female CRD patients had IBS. 29.41% of all CRD patients (13.73% males and 15.69% females) had CKD. 88.24% of all chronic respiratory tract individuals (47.06 % males and 41.18% females) had cardiovascular issues. 9.8% of respiratory patients (5.88% males and 3.92% females) had thyroid. 58.33% of thyroid individuals had diabetes. 25% of thyroid diabetic patients were males, and 33.33% were females. 58.33% of thyroid patients (25% males and 33.33% females) had chronic kidney diseases. 41.67% of thyroid patients (25% males and 16.67% females) had CRD. Surprisingly, all thyroid patients (58.33% males and 41.67% females) had heart issues. This could lead to more research on the association between thyroid and heart disease.

**Co-morbidity prediction analysis using Machine Learning:**

Nowadays, machine learning is used severely to analyze medical reports in the health sector. Machine learning approaches are being applied from disease forecasting [37] to disease prevalence study. In this study, we used Random Forest algorithm to predict the number of disease prevalence by a patient using patients' demographic and physical information. Our task was to classify the patients in 3 classes based on the number of diseases: one, two, more than two. We divided the dataset into 80-20 train-test-split and achieved 68.3% accuracy using random forest and 69.7% accuracy using XGBoost algorithm. According to random forest feature importance, patients' age, weight, and waist -hip ratio have the highest effect on patients' numbers of diseases. On the other hand, gender has the lowest effect, which is shown in Fig. 7.

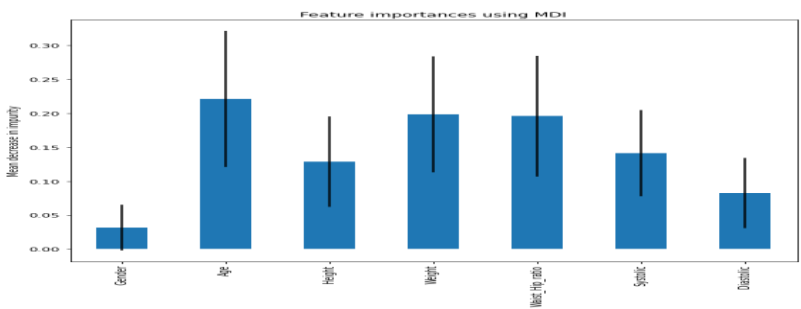

Figure 7. Random Forest Feature Importance in Number of Disease Prediction

**Discussion:**

NCDs have now become a number one threat to modern civilization. These diseases, also known as chronic diseases, are characterized by their long duration and gradual progression. The incidence of cardiovascular diseases, diabetes, cancer, and chronic respiratory diseases is increasing at an alarming rate. Based on the statistical analysis of the participant's demographic information, vital parameters, and disease distribution, this study describes the prevalence of NCDs and the correlation of different NCDs over various age groups of male and female patients in a densely populated region of Dhaka, Bangladesh. Male participants are more affected than females of all diseases studied in this research. 90% of all patients are equal to or above 35 years old, 46% were 35-65 years, and 44% were above 65 (Table 2). This means NCDs in Bangladesh are not only affecting elderly individuals but also are affecting middle-aged people. This is very alarming as 68.36% of the total population of Bangladesh is between 15-64 years old [38]. Bangladesh has the most working-age population, and the country will suffer tremendously if they are affected. Lifestyle factors, such as poor diets, consumption of unhealthy foods, lack of physical labor, occupational stress, and tobacco use can be some of the risk factors for NCDs among middle-aged individuals. All of these might increase the prevalence of NCDs, mainly, CVD and DM. Eventually, younger individuals of 35-45 affected by any NCDs will undergo costly treatment, and if the individual is a sole bread runner in the family, that can lead to a lifelong economic impact. This can have a significant impact on the health and well-being of the patients, thereby, leading to reduced quality of life, disability, and premature death. For elderly population over 65 years, NCDs are the major health concern. 5.33% of Bangladeshi people are 65 and above [38]. In our study, 44% of patients were 65 and above. Often associated with the aging-related degenerative consequences, such as, limited movement, fluctuation of metabolic profiles, loss of cardiovascular functions, bone fragility, loss of cognitive response, and weaken immune system, elder people are more prone to NCDs such as CVD, DM, cancer and CRDs. The healthcare cost could lead to a large financial burden. Prevention and management of NCDs in the elderly population is essential to improve health outcomes and reduce healthcare costs. This includes promoting healthy lifestyles, such as physical activity and healthy eating habits, as well as regular health screenings and management of existing NCDs through medication and lifestyle changes.

Table 2: NCDs between Various Age Group

| Age Group | Gender | DM (%) n = 80 | CKD (%) n = 42 | Heart Issue (%) n = 122 | RTI (%) n = 51 | IBS (%) n = 2 | Thyroid (%) n = 12 |
|---|---|---|---|---|---|---|---|
| <=35 | Male | 1.25 | 4.76 | 6.56 | 1.96 | 0 | 8.33 |
| | Female | 3.75 | 0 | 1.64 | 3.92 | 50 | 0 |
| 35<>65 | Male | 25 | 14.29 | 27.05 | 31.37 | 0 | 16.67 |
| | Female | 21.25 | 14.29 | 17.21 | 19.61 | 0 | 33.33 |
| >=65 | Male | 26.25 | 35.71 | 28.69 | 19.61 | 0 | 33.33 |
| | Female | 22.50 | 30.95 | 18.85 | 23.53 | 50 | 8.33 |

The waist-hip ratio has been considered a better predictor of overall body fitness than body mass index. Greater weight and fat around the waist can be a risk factor for NCDs, such as cardiovascular disease (CVD) and DM. According to the WHO, 0.9 or less in men and 0.85 or less in women can be considered a moderate WHR. In our study, based on the WHR data, every one in five patients was obese (19.17%). This data suggests that Bangladesh is within the non-obese category compared to the US, where the obesity percentage for 2022 was 41.9% [39]. We have observed that there is a significant relation between waist-to hip ratio and CKD (p value 0.042). A relationship was found between waist-to hip ratio and CRD for male patients above 65 (p value 0.04).

We have performed a statistical analysis of blood pressure to observe the frequency of high blood pressure. Elevated blood pressure often does not contain any symptoms and goes unnoticed unless it is monitored regularly. This silent hypertension leads to sudden cardiac arrest, or stroke. We have found that the mean systolic blood pressure lies within the normal range. In the two tailed t test for sample systolic pressure, there was no difference between sample mean and population mean (120 mmHg) (P value 0.229). However, the two tailed t test shows a significant difference between sample mean and population mean (80 mmHg) in diastolic BP measurement. We used a one tailed T test and found the diastolic BP range lies within 75-78 mmHg. Therefore, the diastolic blood pressure of the sample is slightly lower than the average population mean (80 mmHg) (Figure 8).

Diabetes is now called the 'other pandemic' and one of the main growing public health concerns in Bangladesh. The frequency of non-insulin-dependent diabetes mellitus (NIDDM) has been increasing for the last few decades. According to the International Diabetes Federation (IDF), the prevalence of diabetes in Bangladesh is estimated to be around 12.5% of the adult population in 2022, which equates to approximately 13 million people [40]. The rising trend of diabetes in Bangladesh is due to several factors, including an aging population, urbanization, and changes in lifestyle, such as an increase in sedentary behavior and unhealthy diets. Additionally, there is limited access to diabetes screening, treatment, and management in Bangladesh, which contributes to the increasing disease prevalence e burden of diabetes in Bangladesh is significant, as it can lead to various health complications, including cardiovascular disease, blindness, kidney failure, and amputations. Additionally, diabetes can have a significant impact on the quality of life and economic productivity of individuals and the country. In our study, almost half of all hospitalized individuals were diabetic. This disease often co-occurs with one or more diseases, leading to multiple comorbidities. We have seen an age-based progression of this disease. We applied the two-way ANOVA to explore the relationship between multiple variables. The result shows that there is a significant relationship between CVD and CRD in diabetic individuals (p value less than 0.001).

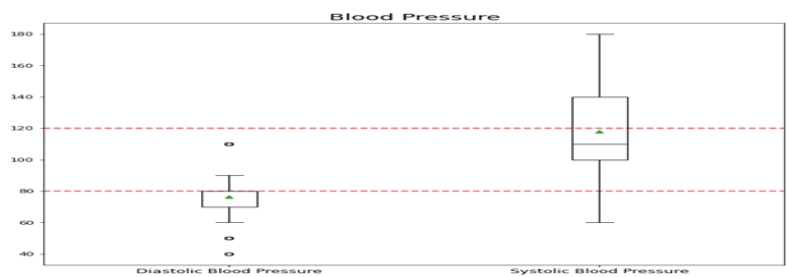

Figure 8: Blood Pressure Sample Mean vs Population Mean.

Chronic kidney failure is one of the leading NCDs, which progressively damages kidney functions. CKD affects both males and females, and the risk factors for CKD include age, hypertension, diabetes, and a family history of kidney disease. 28.77% were CKD patients in our study. Of the total CKD patients, 35% were over 65 years old. This establishes an age-based relationship of CKD. The CKD patients had other coexisting medical conditions. 90% of them had CVDs, which indicates that CKD is a strong predisposing factor to develop cardiovascular disease. More than two-thirds of CKD patients had diabetes mellitus, too.

Irritable bowel syndrome (IBS) is a common gastrointestinal disorder with unknown etiology. The prevalence of IBS in rural Bangladesh is found 24.4% [41]. Often triggered by specific foods, stress, and emotional trauma, IBS can lead to symptoms ranging from bloating, stomach discomfort to frequent bowel movements and unusual abdominal cramps. The burden of IBS in Bangladesh is significant, as it can impact the quality of life, work productivity, and overall well-being of affected individuals. We found only 2 female cases of IBS in our study. One of them had other comorbidities such as DM, CVD and CRD. Data limitation was a barrier for further study in IBS patients. However, the leading NCD among the patients was cardiovascular diseases (83.56% of all patients). It is mainly caused by an unhealthy sedentary lifestyle habit, high blood pressure, high blood cholesterol, obesity, and diabetes. CVD was found to affect more males (62.3%) than females (37.7%). The most possible explanation could be that men are the sole income source for most of the Bangladeshi families, therefore, they are more vulnerable to occupational stress compared to their female counterpart. CVDs are more prone to when associated with other diseases [42]. Moreover, men are more likely to engage in unhealthy lifestyle behaviors such as tobacco use, heavy alcohol consumption, and physical inactivity, which are major risk factors for CVD. Early detection and management of risk factors, such as high blood pressure and cholesterol, can help prevent or delay the onset of CVD. We also saw the onset of thyroid dysfunction among our participants. Thyroid dysfunction can be both hypo and hyperthyroidism, which affects the performance of the thyroid gland. Primarily affecting women, thyroid issues can affect men, too. We found 8.22% of patients with thyroid disorders. Of them, 58.33% had CKD, and all had heart-related issues. We can say that thyroid disorder increases the probability of developing CVDs.

Table 3: P Values Between All Pair of NCDs based on Gender and Age-Group

| Disease | Variables | | DM | CKD | Heart Issue | RTI | Thyroid |
|---|---|---|---|---|---|---|---|
| DM | All | All | - | 0.089 | 0.627 | 0.934 | 0.967 |
| | | Male | - | 0.402 | 0.447 | 0.859 | 0.946 |
| | | Female | - | 0.232 | 0.860 | 0.616 | 0.721 |
| | <=35 | All | - | 0.657 | 0.118 | 0.958 | 0.822 |
|  |  | Male | - | 0.826 | 1.0 | 0.921 | 0.921 |
|  |  | Female | - | 1.0 | 0.349 | 0.971 | 1.0 |
|  | 35<>65 | All | - | 0.971 | 0.7646 | 0.238 | 0.348 |
|  |  | Male | - | 1.0 | 0.917 | 0.434 | 0.349 |
|  |  | Female | - | 0.977 | 0.757 | 0.564 | 0.864 |
|  | >=65 | All | - | 0.038 | 0.346 | 0.152 | 0.606 |
|  |  | Male | - | 0.194 | 0.133 | 0.940 | 0.436 |
|  |  | Female | - | 0.235 | 0.994 | 0.071 | 0.793 |
| CKD | All | All | 0.089 | - | 0.358 | 0.992 | 0.061 |
|  |  | Male | 0.402 | - | 0.969 | 0.993 | 0.603 |
|  |  | Female | 0.232 | - | 0.079 | 0.974 | 0.052 |
|  | <=35 | All | 0.657 | - | 0.561 | 0.522 | 0.03 |
|  |  | Male | 0.826 | - | 1.0 | 0.18 | 0.18 |
|  |  | Female | 1.0 | - | 1.0 | 1.0 | 1.0 |
|  | 35<>65 | All | 0.971 | - | 0.965 | 0.975 | 0.004 |
|  |  | Male | 1.0 | - | 0.541 | 0.863 | 0.363 |
|  |  | Female | 0.977 | - | 0.332 | 0.977 | 0.022 |
|  | >=65 | All | 0.038 | - | 0.864 | 0.946 | 0.984 |
|  |  | Male | 0.194 | - | 0.974 | 0.774 | 0.822 |
|  |  | Female | 0.235 | - | 0.828 | 1.0 | 0.594 |
| Heart Issue | All | All | 0.627 | 0.358 | - | 0.536 | 0.276 |
|  |  | Male | 0.447 | 0.969 | - | 0.994 | 0.605 |
|  |  | Female | 0.86 | 0.079 | - | 0.269 | 0.436 |
|  | <=35 | All | 0.118 | 0.561 | - | 1.0 | 0.765 |
|  |  | Male | 1.0 | 1.0 | - | 1.0 | 1.0 |
|  |  | Female | 0.349 | 1.0 | - | 0.729 | 1.0 |
|  | 35<>65 | All | 0.764 | 0.965 | - | 0.803 | 0.452 |
|  |  | Male | 0.917 | 0.541 | - | 0.793 | 0.799 |
|  |  | Female | 0.757 | 0.332 | - | 0.977 | 0.511 |
|  | >=65 | All | 0.346 | 0.864 | - | 0.631 | 0.755 |
|  |  | Male | 0.133 | 0.974 | - | 0.959 | 0.825 |
|  |  | Female | 0.994 | 0.828 | - | 0.233 | 0.934 |
| RTI | All | All | 0.934 | 0.992 | 0.536 | - | 0.877 |
|  |  | Male | 0.859 | 0.993 | 0.994 | - | 0.792 |
|  |  | Female | 0.616 | 0.974 | 0.269 | - | 1.0 |
|  | <=35 | All | 0.958 | 0.522 | 1.0 | - | 0.117 |
|  |  | Male | 0.921 | 0.18 | 1.0 | - | 0.018 |
|  |  | Female | 0.971 | 1.0 | 0.729 | - | 1.0 |
|  | 35<>65 | All | 0.238 | 0.975 | 0.803 | - | 0.506 |
|  |  | Male | 0.434 | 0.863 | 0.793 | - | 0.495 |
|  |  | Female | 0.564 | 0.977 | 0.977 | - | 0.864 |
|  | >=65 | All | 0.152 | 0.946 | 0.631 | - | 0.454 |
|  |  | Male | 0.94 | 0.774 | 0.959 | - | 0.523 |
|  |  | Female | 0.071 | 1.0 | 0.233 | - | 0.545 |
| Thyroid | All | All | 0.967 | 0.061 | 0.276 | 0.877 | - |
|  |  | Male | 0.946 | 0.603 | 0.605 | 0.792 | - |
|  |  | Female | 0.721 | 0.052 | 0.436 | 1.0 | - |
|  | <=35 | All | 0.822 | 0.03 | 0.765 | 0.117 | - |
|  |  | Male | 0.921 | 0.18 | 1.0 | 0.018 | - |
|  |  | Female | 1.0 | 1.0 | 1.0 | 1.0 | - |
|  | 35<>65 | All | 0.348 | 0.004 | 0.452 | 0.506 | - |

|  |  | Male | 0.349 | 0.363 | 0.799 | 0.495 | - |
|  |  | Female | 0.864 | 0.022 | 0.511 | 0.864 | - |
|  | >=65 | All | 0.606 | 0.984 | 0.755 | 0.454 | - |
|  |  | Male | 0.436 | 0.822 | 0.825 | 0.523 | - |
|  |  | Female | 0.793 | 0.594 | 0.934 | 0.545 | - |

Chronic lung diseases, including chronic obstructive pulmonary disease (COPD) and asthma, are one of the top 10 NCDs responsible for 3.9 million deaths each year [42]. Air pollution is a significant contributor to the high rates of chronic respiratory disease in Bangladesh. Bangladesh is one of the most polluted countries in the world, with air pollution levels that are among the highest in the region [44]. This air pollution is largely due to emissions from industrial activities, vehicular traffic, and the burning of solid fuels for cooking and heating. Exposure to high levels of air pollution, mainly particulate matter, can increase the risk of chronic respiratory diseases, lung cancer, heart disease, and other health problems. Our study showed that 34.93% of total patients had chronic respiratory diseases. Of them, 32.5% were in the 35-65 years age range. Almost half of them had diabetes, and one-third had CKDs.

Bangladesh is the eighth most populated country in the world. Dhaka has the highest population density of 30,093 residents per square kilometer. This overpopulation is creating social, economic, and environmental issues that are triggering NCD incidences gradually. The effect of urbanization on NCDs is very evident. The urbanization process amplified the lifestyle risk factors for NCDs: unhealthy diets, tobacco use, harmful alcohol intake, and physical inactivity [45]. As a result, it increases the average BMI such that the least urbanized countries' BMI average is 2.3 kg/m2 lower than most urbanized ones, and the cholesterol difference was found to be 0.40 mmol/L. At the same time, the least urbanized countries are expected to have an up to 3.2 p.p. lower prevalence of diabetics among women, according to a study from 1980-2008 [46]. Socioeconomic status also has a vital role in NCDs. Individuals in higher socio-economic group living in urban setting have higher prevalence of diabetics, hypertension, obesity than non-manual workers [47]. In Bangladesh according to Bangladesh Demographic and Health Survey (BDHS 2017–18), wealth status contributed approximately 25.71% and 43.41% of total inequality in hypertension and diabetics respectively [48].

In our study, we observed the association of NCDs (DM, CKD, IBS, CVD, CRD, Thyroid) between gender, age groups, and abdominal adiposity. We have observed that most of these chronic diseases co-exist with other NCDs. For instance, 31.5% of the population has only one NCD, 30.1% has two NCDs, and 38.3% has more than two NCDs. Here we can see an interdependency of chronic diseases. We also performed t-tests for different risk factor variables. We found a relationship between CKD and thyroid under all age groups (p-value 0.061). We also observed age and gender-based statistical analysis of different NCDs. Males under 35 years have a statistically significant relationship between thyroid and chronic respiratory diseases (p-value 0.018). Moreover, we found a relation between DM and CKD among patients over 65. There has been a significant relation between CKD and Thyroid ($P < 0.05$) for both below 35 years and 35-65 years age group. The overall association between each pair of NCDs in each category is listed in Table 3 with their respective P-values. We also did a two-way ANOVA test and found that two independent diseases, heart issues and chronic respiratory illness, in combination, affect diabetes. Moreover, the combination of DM and RTI also affected the CKD in male patients over 65 years old. We were able to detect comorbidity using tree-based machine learning algorithms. XGBoost gave the highest accuracy 69.7% with low recall and precision score. Using the random forest feature importance, we learned that comorbidity depends on several physical conditions like age, weight, waist hip ratio. Among them, age is the most important factor and most of the age > 65 patients affected by more than two NCDs.

## Limitations:

Dhaka and the rural areas have different socio-economic conditions, environmental pollution, and air index profiling. Although we included patients of both areas and socio-economic conditions in this study, more data would have been more significant in determining the NCDs prevalence in Bangladesh. A longitudinal population-based study, therefore, can infer more interpretation about the year-wise pattern of NCDs in Bangladesh. However, some NCDs like IBS and thyroid had very little data, so we could not interpret the analysis and comorbidity associated with this information. Moreover, due to patient confidentiality, we could not collect more information about patient demographic and habits which restricts us from drilling down more specific causes related to individual NCDs.

## Conclusion:

NCDs are the most significant challenge of this 21$^{st}$ century. During the covid-19 era, NCD risk factors made people more vulnerable to covid-19, and 75% of countries reported interruptions to NCD services [49]. Therefore, in this post covid era, we are watching a sudden increase in NCDs in the world's population. Moreover, with urbanization and industrialization over the last 30-50 years, the need for rigorous physical activity for both genders has also lessened. At the same time, unhealthy diets, tobacco use, and alcohol consumption are increasing. All these factors increase the risk of NCDs even in the younger population. Therefore, we need urgent policies to control NCDs' prevalence. This study can be vital in policy making as it represents the prevalence of NCDs based on the population's demographic and physical information and finds the most vulnerable groups. It also explored the associations between NCDs, which will help healthcare practitioners understand comorbidities, NCD screening, and palliative care.

**Funding:**

No funding was used to write this research paper.

**Conflict of Interest:**

The authors declare that they do not have any conflicts of interest.

**Acknowledgement**

We acknowledge that this study has been made public previously on a preprint server [50].